\documentclass[twocolumn,showpacs]{revtex4}

\usepackage{graphicx}%
\usepackage{dcolumn}
\usepackage{amsmath}

\begin{document}

\preprint{gr-qc/0106010}

\title{Support for a prosaic explanation for the anomalous
acceleration of Pioneer 10 and 11}

\author{Louis K. Scheffer}
 \email{lou@cadence.com}
\affiliation{%
Cadence Design Systems \\
555 River Oaks Parkway \\
San Jose, CA 95134
}%

\date{\today}

\begin{abstract}
Anderson, {\it et al.}\ find the measured trajectories of Pioneer 10 and 11 
spacecraft deviate from the trajectories computed from known forces acting
on them.  This unmodelled acceleration can be accounted for by
non-isotropic radiation of spacecraft heat. This explanation was first
proposed by Murphy, but Anderson, {\it et al.}\ felt it could not explain
the observed effect.  This paper includes new calculations on the expected 
magnitude of this effect, based on the relative emissivities of the
different sides of the spacecraft, as estimated from the known
spacecraft construction.  The calculations indicate
the proposed effect can account for most, if not all, of the unmodelled
acceleration.

\end{abstract}

\maketitle

\section{INTRODUCTION}
\label{intro}

In \cite{anderson}, Anderson {\it et al.}\ compare the measured trajectory
of spacecraft against the theoretical trajectory  computed from known
forces acting on the spacecraft.  The find a small but significant
discrepancy, referred to as the unmodelled or anomalous acceleration.
It has an approximate magnitude of $\rm 8\times10^{-8}\;cm\;s^{-2}$ 
directed approximately towards the Sun.  Needless to say, {\it any} 
acceleration
of {\it any}  object that cannot be explained by conventional physics is
of considerable interest, and these spacecraft have been tracked very
accurately over a period of many years, so the data is quite reliable.
Explanations for the acceleration fall into two general categories - either
new physics is needed or some conventional force has been overlooked.

In \cite{murphy}, Murphy proposes that the source of the acceleration 
is non-isotropic radiation from the main spacecraft
bus.  In \cite{usmurphy} and \cite{anderson01}, Anderson {\it et al.}\ reply 
with reasons that lead them to believe
the proposed mechanism is not correct.  Their main argument is that the
cooling louvers are known to be closed, as opposed to open louvers assumed by
Murphy.  This, they felt, would result in isotropic radiation of heat.

This paper argues that non-isotropic radiation is in fact the most likely 
cause for the unmodelled acceleration, and the objections to it
are overstated.  Put most simply, all the heat dissipated in the body of the
spacecraft must eventually be radiated by some externally visible surface.
Assuming a uniform internal temperature, the power emitted from each 
surface is proportional to the area times the effective
emissivity of the surface.   Using the known construction and 
temperatures of Pioneer 10/11, we conclude most of the heat will
be radiated from the anti-sunward faces of the spacecraft, even though
the louvers are closed.  Rough numerical calculations indicate this
effect accounts for much, it not all, of the anomalous acceleration.

\section {PREVIOUS WORK}
\label{pioneer}

For the convenience of the reader, 
section 2.1 consists of direct quotes from \cite{anderson01}, covering 
the relevant details of the Pioneer spacecraft.  The references cited in
this section are included in the references of this paper.  Many other
web and paper descriptions are available \cite{piodoc,pioweb}.

In section 2.2 we summarize the existing literature on the hypothesis that 
non-isotropic radiation from the main spaecraft bus is responsible for the 
unmodelled acceleration.


\subsection{General description of the Pioneer spacecraft, 
from \cite{anderson01}}
\label{sec:pio_description}

The main equipment compartment is 36 cm deep.  The hexagonal flat top and
bottom have 71 cm long sides.  The equipment compartment  provides a
thermally controlled environment for scientific instruments.
  
The majority of the
spacecraft electrical assemblies are located in the central hexagonal
portion of the compartment, surrounding a 16.5-inch-diameter
spherical hydrazine tank.  Most of the scientific instruments' electronic
units and internally-mounted sensors are in an instrument bay (``squashed''
hexagon) mounted on one side of the central hexagon. The equipment
compartment is contained within a structure of aluminum honeycomb which
provides support and meteoroid protection. It is  covered with insulation
which, together with louvers under the mounting platform, provides 
passive thermal control.

The spacecraft instrument compartment is thermally controlled between 
approximately $0$ F and 90 F. This is done with the aid  of
thermo-responsive louvers located at the bottom of the  equipment
compartment. These louvers are adjusted by bi-metallic springs.    They
are completely closed below $\sim40$ F and completely  open above 
$\sim 85$ F.  This  allows controlled heat to escape in the
equipment compartment.  Equipment is kept within an operational range of
temperatures by multi-layered blankets of insulating aluminum plastic. 
Heat is provided by electric heaters, the heat from the instruments 
themselves,  and by twelve one-watt radioisotope heaters powered directly
by non-fissionable plutonium 

The essential platform temperature as of the year 2000 is still within
acceptable limits at $-41$ F; the nominal range is between $-63$ F and
180 F. The RF power output from the TWT-A traveling-wave-tube amplifier
is still within normal parameters, having a value of 36 dBm\cite{dBm}.
(The nominal range is 27 to 40 dBm.) 

The spacecraft needs 100 W to power all systems, including 26 W for the
science instruments.  Previously, when the available electrical power was
greater than 100 W,  the excess power was either thermally radiated into
space by a  shunt-resistor radiator or it was used to charge a battery in
the equipment compartment. 

At present only about 65 W of power is available to Pioneer 10
\cite{theorypower}.  Therefore,  all the instruments are no longer able to
operate simultaneously.   But the power subsystem  continues to provide
sufficient power to support the current spacecraft load: transmitter,
receiver, command and data handling, and the Geiger Tube Telescope  (GTT)
science instrument.



\subsection{Non-isotropic radiative cooling of the spacecraft}
\label{subsec:mainbus}

Murphy suggests that the  anomalous acceleration seen in the
Pioneer 10/11 spacecraft can be, ``explained, at least in part, by
non-isotropic  radiative cooling of the spacecraft.'' \cite{murphy}
Anderson, {\it et al.}\ argue in reply that this explanation is flawed, since
``by nine AU the
actuator spring  temperature had already reached $\sim$40  F
\cite{piodoc}.  This means the  louver doors were   closed  (i.e.,  the
louver angle was zero)  from where we obtained our data. 
Thus, from that time on of the radiation properties,  the contribution
of the thermal radiation to the Pioneer anomalous acceleration should be 
small.''  They also argue that the spacecraft power is decreasing, but
the unmodelled acceleration is not.

\section{Discussion}
The sunward side of the spacecraft is the back, and the anti-sunward side,
in the
direction of motion, is the front\cite{rearfront}.  We consider thermal
radiation from the spaecraft with the louvers closed, as they have been
since 9 AU.
We consider the radiation from the front, back, and sides of the
spacecraft bus.  Assuming the compartment is at a constant temperature,
the radiation from each surface will be determined by the effective
emissivity of that surface times its area.

First, from the known sizes the front and back of the central equipment
compartment have about 1.3 m$^2$ area, and the sides about 1.5 m$^2$ total.
The sides (and presumably the rear) of the compartment are covered with
multi-layer insulation.  From \cite{MLI}, multilayer insulation from 
spacecraft typically has an effective emissivity of 0.002 to 0.02.
Assuming a value of 0.01, and an internal temperature of 233 K, the
instrument compartment will lose about 5 watts through the sides and
back.  Ignoring conduction losses through connecting wires and struts,
then the rest of the power (about 59 watts as of 1998) must be radiated 
from the front.

Is it reasonable for the front to radiate this much?  At 233 K, the area
times the emissivity must equal 0.35 m$^2$.  If the surface was flat, this
would require an average emissivity of 0.27.  From a picture of the 
Pioneer 10 replica in the National Air and Space Museum \cite{NASM}, 
the front of the spacecraft is rather complex, with supports, louvers, 
and a variety of surface finishes . A composite emissivity of 0.27 
seems reasonable.

The main conclusion seems quite robust.  Multi-layer insulation is
specifically designed to reduce heat losses, whereas the louvers have at
most one layer of obstruction even when closed. (The Rosetta louvers, for
example, have an emissivity range of 0.09-0.76\cite{Rosetta}.)  Therefore
a majority of the heat will be radiated from the front of the spacecraft.

\subsection{Radio beam power}

The radio beam power is reported in \cite{anderson01} two different ways, as 8 watts and
36 dBm.  These values are not consistent, since 36 dBm $\sim$ 3.98 watts.
Assuming the dBm figure is correct, the smaller
value of radiated power reduces the value of the anomalous acceleration.

\subsection{Effect on the unmodelled acceleration}

Anderson reports an unmodelled acceleration  for Pioneer 10 of 
$7.84 \times 10^{-8}$ cm/sec$^2$.  If the radio
beam is 4 watts instead of 8, this becomes $7.31\times 10^{-8}$ cm/sec$^2$.
As of 1998, the power in the main bus was about 68 watts.  
If 4 watts goes into the radio beam, and 5 watts through the sides of the
instrument compartment, then 59 watts must radiate from the front.
If this radiated as if from a flat plate\cite{murphy}, then this accounts for
$5.2 \times 10^{-8}$ cm/sec$^2$.  If it is radiated as a collimated beam, 
then the resulting acceleration is $7.8 \times 10^{-8}$ cm/sec$^2$. 
The true result should lie somewhere between these two extremes.

This explanation also explains some other puzzles: the values of acceleration
of Pioneer 10 and 11 would be expected to be similar, but not identical,
as observed. The acceleration would not have a strong effect on the spin;
since the louvers are closed, the radiation will generate little torque.
Other spacecraft, built along the same general principles,
would be expected to show a similar effect, but planets and other large
bodies would not, as is observed.

\subsection{Why is the acceleration not dropping as the power level drops?}

This is covered in more detail by Murphy\cite{murphy}.
As the total power level drops, some parts of the
spacecraft will dissipate less power (such as the shunt
regulators and science instruments) and some will remain the
same (command and control, and the transmitter, for example).  
If the hypothesis put forth in this paper is true, then the acceleration
of the spacecraft should be proportional to the power dissipated in the
central equipment compartment.
The construction of the spacecraft puts most of the experiments on the
outside and most of the housekeeping functions in the central compartment.
Since the central compartment contains mostly systems essential 
to the operation of the spacecraft, the power dissipated within it 
has changed comparatively little.  This explains why the unmodelled
acceleration has changed little despite the considerable reduction in
spacecraft power.

\section{Conclusions and future works}

There is surely an unmodelled effect on the Pioneer spacecraft, based only
on its thermal characteristics.  Rough estimates show
it may account for most, if not all, of the unmodelled acceleration.

More detailed modeling, using the Pioneer materials, construction details,
and history, could provide a much better estimate of the magnitude of this
effect.  A suitably detailed thermal model, measured in a cold vacuum
chamber, would provide the strongest evidence for or against this hypothesis.

Proposed missions such as LISA\cite{LISA} will attempt to detect
gravitational waves by measuring the changes in distance between spacecraft
about $5\times 10^{9}$ meters apart with an accuracy of a few picometers. 
They will then look for unmodelled displacement.  To find the anticipated 
small effects, the LISA project isolates the
spacecraft from non-gravitational disturbances by actively controlling them
to keep them centered around a ``drag-free'' proof mass.  
This technique should keep the
acceleration induced by non-gravitational forces to about 
$10^{-13}$ cm/sec$^2$, or about 1 part in $10^5$ of the proposed anomalous
acceleration.  Given this accuracy, and the proposed formation of a
triangle inclined to the ecliptic, an anomalous
acceleration as proposed for Pioneer 10/11 will be
easily distinguishable from the conventional gravitational forces.  If no
anomalous accelerations are detected in this more precise experiment, 
then almost surely the unmodelled acceleration of Pioneer 10 and 11 
is caused by some overlooked prosaic source such as the one proposed here.

%




\begin{thebibliography}{99}
\bibitem{dBm}  dBm is used by radio engineers as a measure of received
power. It stands for decibels referred to a milliwatt.  [Note: the articles
by Anderson define it as power below a milliwatt, which is inconsistent
with the sign of their usage.]
\bibitem{theorypower}
This is a ``theoretical value,'' which does not account for inverter
losses, line losses, and such. It is interesting to note
that at mission acceptance, the total ``theoretical'' power was 175 Watts.
\end{thebibliography}
\end{document}